\documentclass[]{interact}

\usepackage{epstopdf}
\usepackage[caption=false]{subfig}

\usepackage[numbers,sort&compress]{natbib}
\bibpunct[, ]{[}{]}{,}{n}{,}{,}

\theoremstyle{plain}

\theoremstyle{definition}

\theoremstyle{remark}

\begin{document}


\title{Central-moment-based discrete Boltzmann modeling of compressible flows}

\author{
\name{Chuandong Lin\textsuperscript{a,b,c}, Xianli Su\textsuperscript{a,d}, Linlin Fei\textsuperscript{e} and Kai H. Luo\textsuperscript{b,d} \thanks{CONTACT Chuandong Lin Email: linchd3@mail.sysu.edu.cn; Xianli Su Email: suxli@mail2.sysu.edu.cn; Linlin Fei Email: linfei@ethz.ch; Kai Hong Luo Email: k.luo@ucl.ac.uk}}
\affil{\textsuperscript{a}Sino-French Institute of Nuclear Engineering and Technology, Sun Yat-sen University, Zhuhai, 519082, China; \\ \textsuperscript{b}Center for Combustion Energy, Key Laboratory for Thermal Science and Power Engineering of Ministry of Education, Department of Energy and Power Engineering, Tsinghua University, Beijing, 100084, China; \\ \textsuperscript{c}National University of Singapore, Department of Mechanical Engineering, National University of Singapore, 119077, Singapore; \\ \textsuperscript{d}Department of Mechanical Engineering, University College London, London, WC1E 7JE, United Kingdom; \\ \textsuperscript{e}Chair of Building Physics, Department of Mechanical and Process Engineering, ETH Z\"urich, 8092, Switzerland}
}

\maketitle

\begin{abstract}
In this work, a central-moment-based discrete Boltzmann method (CDBM) is constructed for fluid flows with variable specific heat ratios. The central kinetic moments are employed to calculate the equilibrium discrete velocity distribution function in the CDBM. In comparison to previous incompressible central-moment-based lattice Boltzmann method, the CDBM possesses the capability of investigating compressible flows with thermodynamic nonequilibrium effects beyond conventional hydrodynamic models. Unlike all existing DBMs which are constructed in raw-moment space, the CDBM stands out by directly providing the nonequilibrium effects related to the thermal fluctuation. The proposed method has been rigorously validated using benchmarks of the Sod shock tube, Lax shock tube, shock wave phenomena, two-dimensional sound wave, and the Taylor-Green vortex flow. The numerical results exhibit an exceptional agreement with theoretical predictions.
\end{abstract}


\section{Introduction}\label{SecI}

Compressible flows, characterized by both hydrodynamic and thermodynamic nonequilibrium effects, are ubiquitous in natural and engineering contexts, including applications such as inertial confinement fusion, gas pipelines, jet engines, and rocket motors \cite{balachandran2006}. Notably, modern high-speed airplanes and the jet engines that power them are wonderful examples of the application of compressible flows. Additionally, during the reentry of an aircraft into the atmosphere at high Mach numbers, the surrounding shock wave induces significant acceleration and heating of the air. Therefore, in the modern world of aerospace and mechanical engineering, a deep understanding of the principles of compressible flows is essential \cite{anderson2021modern}.

To predict the thermal process and nonequilibrium effects, there are three main categories of numerical methodologies. The first class involves modifying traditional macroscopic models, for example, the Navier-Stokes (NS) equations plus slip boundary conditions \cite{aliabadi1993space,hughes2010jsc,aoki2017josp}, the NS-type equations with multi-components and/or multi-temperatures \cite{pedro2019numerical,shershnev2022numerical}, Burnett and super-Burnett equations \cite{agarwal2002burnett,jadhav2017force}, etc. Since the macroscopic models are constructed under the continuum assumption, the application scope is limited at small Knudsen numbers. The second type encompasses microscopic models, such as the molecular dynamics (MD) \cite{sun1992pra,kandemir2012molecular}. The MD provides details of physical systems, but is only applicable to small temporal and spatial domains due to an excessive computational burden. 

Moreover, to address the aforementioned challenges, the third category of methods encompasses mesoscopic approaches based on kinetic theory, such as the direct simulation Monte Carlo (DSMC) method \cite{Bird1994}, discrete velocity methods (DVM) \cite{Mieussens2000JCP,YLM2017CNF}, (discrete) unified gas-kinetic schemes ((D)UGKS) \cite{XK2010JCP,CSZ2012JCP,GZL2015PRE,GZL2021AIA}, the lattice Boltzmann method (LBM) \cite{SucciBook,GuoBook2013}, and the discrete Boltzmann method (DBM) \cite{Xu2012FP,YB2013fop}, among others. These approaches act as a bridge connecting microscopic and macroscopic scales. The DSMC method, first introduced by G. A. Bird, has been extensively developed and widely applied in rarefied gas dynamics, particularly for high-speed flows \cite{Bird1994}. However, DSMC is not well-suited to the simulation of low-speed flows, and has to balance the noise level and computational efficiency \cite{Hadjiconstantinou2003JCP}. DVM represents another widely used approach, where the continuous particle velocity space is discretized into a finite set of velocity coordinate points, and numerical quadrature is employed to approximate the integration of moments \cite{Mieussens2000JCP,YLM2017CNF}. However, for high-speed compressible flows, especially in the near continuous flow region, the DVM exhibits limited computational efficiency \cite{YLM2017CNF}. Besides, in order to solve the discrete velocity Boltzmann equation, Xu et al. \cite{XK2010JCP,CSZ2012JCP} and Guo et al. \cite{GZL2015PRE,GZL2021AIA} proposed the unified gas-kinetic scheme (UGKS) and the discrete unified gas-kinetic scheme (DUGKS), respectively. In UGKS, the local integration of the discrete velocity Boltzmann equation is used to compute the flux of the distribution function, whereas in DUGKS, the flux is derived directly from the evolution equation.

Among the kinetic methodologies, the LBM stands out \cite{SucciBook,GuoBook2013}. The LBM has emerged as a competitive scheme for simulating various complex flows due to its canonical ``collision-streaming" algorithm which disentangles non-linearity and non-locality and is easy to implement. Owing to these inherent advantages, LBM has been successfully applied to simulate various physical problems including multiphase \cite{haghani2018PRE}, reactive \cite{di2012lattice}, magnetohydrodynamic \cite{cherkaoui2022magnetohydrodynamic}, nano- \cite{afzalabadi2020lattice}, biomechanics \cite{xu2021study}, and porous media flow \cite{LX2022IJTS}. Although the LBM has achieved significant success in simulating nearly incompressible complex flows, its application to compressible flows continues to present significant challenges.

Alexander et al. devised the first multi-speed lattice Boltzmann model containing 13 discrete velocities, representing the earliest application of the LBM to the compressible NS equations \cite{alexander1993PRE}. Shortly after the introduction of Alexander's model, in 1993, Qian proposed a series of multi-speed models based on the DnQb thermal lattice Bhatnager--Gross--Krook (BGK) isothermal model for thermohydrodynamics, in which a proper internal energy is introduced and the energy equation is obtained \cite{QYH1993JSC}. In 1994, Chen et al. introduced a thermal lattice BGK model capable of recovering the standard compressible NS equations through a higher-order velocity expansion of the Maxwellian-type equilibrium distribution \cite{CY1994PRE}. However, this model was constrained by assumptions of zero bulk viscosity and unit Prandtl number \cite{CY1994PRE}. In 1997, Chen et al. constructed a two time relaxation parameters thermal lattice BGK model to achieve adjustable Prandtl number \cite{CY1997JSC}. In the same year, McNamara et al. derived the distribution functions of thermal LBM using a series of moment conservation equations, enabling variability in the Prandtl number within this model \cite{Mcnamara1997JSP}. Subsequently, compressible lattice Boltzmann models with a flexible specific heat ratio were introduced by Hu et al. in 1997 \cite{Hu1997ACTA} and Yan et al. in 1999 \cite{Yan1999PRE}. With the aim of simulating compressible flows, Sun et al. developed an adaptive compressible LBM based on a simple delta function, where the lattice velocities vary with mean flow velocity and internal energy \cite{Sun2000PRE,Sun2000JCP,Sun2000AMS,Sun2003PRE,Sun2004CNF}. In 2007, Li et al. proposed a coupled double-distribution-function LBM for the NS equations with a flexible specific heat ratio and Prandtl number \cite{LQ2007PRE}.

In recent years, Yang et al. developed a platform for constructing one-dimensional compressible LBM and subsequently extended the finite volume LBM to simulate compressible flows, including shock waves \cite{YLM2012AIAMM,YLM2013CNF}. Li et al. introduced a novel LBM model designed for fully compressible flows. Building upon a multi-speed model, an extra potential energy distribution function is introduced to recover the full compressible NS equations, featuring a flexible specific heat ratio and Prandtl number \cite{LK2015IJFNMF}. In 2018, Dorschner et al. constructed a particles-on-demand (PonD) method which is suitable for both incompressible and compressible flows \cite{Dorschner2018PRL}. In the PonD method, the discrete particle velocities are defined relative to a local reference frame, determined by the local flow velocity and temperature, which vary in both space and time. Subsequently, Kallikounis et al. further refined PonD method with a revised reference frame transformation using Grad's projection to enhance stability and accuracy \cite{Kallikounis2022PRE}. In 2024, Kallikounis et al. improved the PonD method for variable Prandtl number by employing the quasi-equilibrium relaxation, and the model was validated via simulations of high Mach compressible flows \cite{Kallikounis2024PRE}. In the same year, Ji et al proposed an Eulerian realization of the PonD for hypersonic compressible flows, where a kinetic model formulated in the comoving reference frame \cite{JY2024JFM}.

Generally, lattice Boltzmann methods (LBMs) are broadly classified into the single-relaxation-time (SRT) model and the multiple-relaxation-time (MRT) model. The SRT or BGK operator is the simplest collision operator in the standard LBM, where all distribution functions relax to their local equilibrium values at a constant rate. However, the BGK-LBM may encounter stability issues in the zero-viscosity limit and/or for non-vanishing Mach numbers \cite{GuoBook2013,Coreixas2019PRE,Coreixas2020PT}. To mitigate this issue, numerous strategies have been proposed, including modifications to the numerical discretization, collision model, or both \cite{Coreixas2019PRE,Coreixas2020PT}. In 2006, Geier et al. introduced a cascaded operator, conducting collisions in the central-moment space rather than that of raw moments as in the MRT-LBM\cite{Geier2006PRE}. Consequently, the corresponding method gradually interpreted as ``central-moment-based" lattice Boltzmann method (CLBM) \cite{lycett2014POF,lycett2014CMP,NY2016IJNMF,FLL2017PRE,FLL2018IJHMT,LX2024ICHMT}. However, the CLBM does not necessarily provide enhanced stability. For example, when relaxation in the central moment space is implemented using a uniform relaxation parameter for all moments, the approach reduces to a BGK collision operator with an extended equilibrium \cite{Asinari2008PRE,DR2019PRE}. To improve the stability of CLBMs, the relaxation frequencies of higher-order moments, including the trace of second-order moments, should be set to one or a value close to one \cite{Geier2017ICP,Wissocq2022JCP}.

In this work, we constructed a central-moment-based discrete Boltzmann method (CDBM) tailored specifically for the simulation of compressible flows. The ``central-moment-based" indicates that physical quantities and higher-order kinetic moments are computed through central moments. Unlike the majority of previously proposed CLBMs, which are constrained to incompressible flows and neglect the thermodynamic nonequilibrium effects inherent to the Boltzmann equation, the CDBM possesses significantly broader applicability. Based upon the Boltzmann equation, in addition to capturing the general hydrodynamic behaviors described by the NS model, the CDBM provides deeper insights into more detailed thermodynamic nonequilibrium behaviors in various complex fluids. Moreover, distinct from all existing DBMs constructed in raw-moment space \cite{Xu2012FP,YB2013fop,LCD2014PRE,Lai2016PRE,GYB2018PRE,SU2020E,chen2021fop,sxl2022ctp,su2023CTP,Gan2015SM,Lin2016CNF,liu2023discrete}, the CDBM, using the peculiar velocity, can quantify the nonequilibrium effects related to the thermal fluctuation directly. These distinctive features make the CDBM particularly valuable for investigating compressible flows that exhibit significant thermodynamic nonequilibrium effects.

\section{Research methodology}\label{SecII}

\begin{figure}
	\begin{center}
		\includegraphics[width=0.55\textwidth]{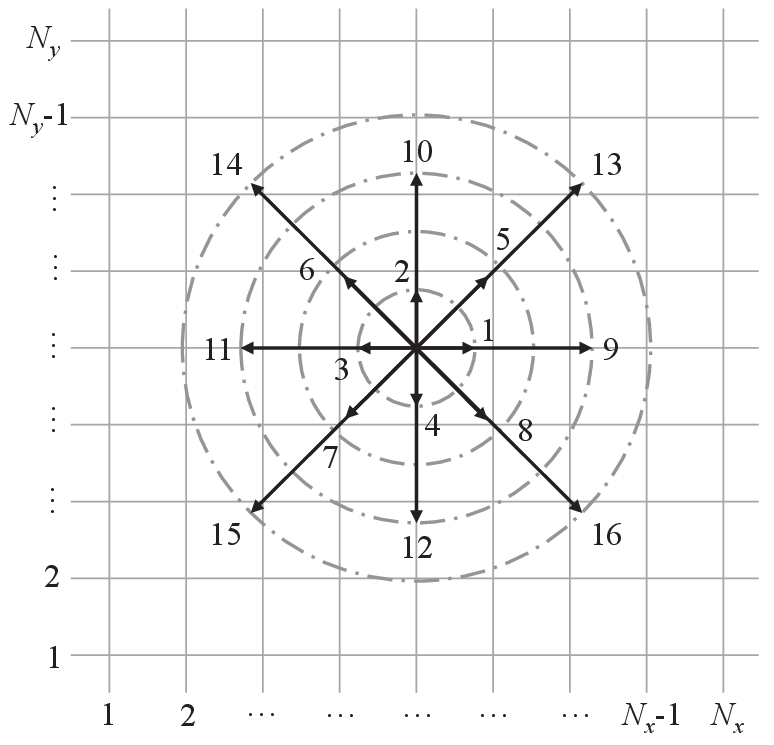}
	\end{center}
	\caption{The schematic of discrete velocities.}
	\label{Fig01}
\end{figure}

The governing equations for the CDBM are expressed as follows: 
\begin{equation}
	\frac{\partial {{f}_{i}}}{\partial t}+{{\mathbf{v}}_{i}}\cdot\frac{\partial {{f}_{i}}}{\partial {{r}_{\alpha }}}=-\frac{1}{\tau } \left( {{{{f}}}_{i}}-{f}_{i}^{eq} \right)
	\label{governing equation}
	\tt{,}
\end{equation}
where $f_i$ indicate the discrete distribution functions, with the subscript $i$ denoting the velocity index. The discrete velocities are given by:
\begin{equation}
	{{\mathbf{v}}_{i}}=\left\{ \begin{array}{*{35}{l}}
		\mathit{{v}_{a}}\left( \pm 1,0 \right), & 1\le i\le 4,  \\
		\mathit{}{{v}_{b}}\left( \pm 1,\pm 1 \right), & 5\le i\le 8,  \\
		\mathit{{v}_{c}}\left( \pm 1,0 \right), & 9\le i\le 12,  \\
		\mathit{}{{v}_{d}}\left( \pm 1,\pm 1 \right), & 13\le i\le 16, 
	\end{array} \right.
\end{equation}	
where $v_a$, $v_b$, $v_c$ and $v_d$ are adjustable. Besides, in order to describe the vibrational and/or rotational energies, $I$ is introduced for extra degrees of freedom due to vibration and/or rotation, and the corresponding parameters for degrees of freedom, ${\eta}_i$, are defined as follows,
\begin{equation}
	{{\eta }_{i}}=\left\{ \begin{array}{*{35}{l}}
		{{\eta }_{a}}, & 1\le i\le 4,  \\
		{{\eta }_{b}}, & 5\le i\le 8,  \\
		{{\eta }_{c}}, & 9\le i\le 12,  \\
		{{\eta }_{d}}, & 13\le i\le 16.  \\
	\end{array} \right.
\end{equation}	
Figure \ref{Fig01} exhibits the schematic of discrete velocities.
It is worth noting that the values of these parameters are chosen based on the specific application to ensure numerical stability. In general, the magnitudes of $v_a$, $v_b$, $v_c$ and $v_d$ should approximate key physical quantities such as flow velocity $\mathbf{u}$, sound speed $v_{s}=\sqrt{\gamma T}$, and shock speed, etc. Similarly, the magnitudes of ${\eta }_{a}$, ${\eta }_{b}$, ${\eta }_{c}$ and ${\eta }_{d}$ should be around the value of $\sqrt{I T}$. This is because, in thermodynamic equilibrium, $m\bar{\eta}^2 /2= IT/2$ where m = 1 is the particle mass, and $\bar{\eta}^2$ represents the average value of $\eta^{2}$, according to the equipartition of energy theorem. Hence, $\bar{\eta}=\sqrt{IT}$, and the values of ${\eta }_{a}$, ${\eta }_{b}$, ${\eta }_{c}$ and ${\eta }_{d}$ should be around $\bar \eta$. 

The relaxation time $\tau$ is a constant which governs the relaxation speed of ${{{f}}_{i}}$ towards ${f}_{i}^{eq}$. ${{{f}}_{i}}$ and ${f}_{i}^{eq}$ represent the discrete velocity distributions and the corresponding equilibrium counterparts, respectively. The equilibrium discrete distribution functions, ${f}_{i}^{eq}$, are uniformly computed via a matrix inversion method \cite{Lin2019PRE}. In contrast to prior DBM constructions, this study employs central kinetic moments. 

To recover the NS equations, through the Chapman-Enskog expansion, the discrete equilibrium distribution functions should satisfy the following seven central kinetic moments,
\begin{equation}
	\int {\int {{f^{eq}}\Psi d{\bf{v}}d\eta } }  = \sum\nolimits_i {f_i^{eq}{\Psi _i}}
	\label{moment}
	\tt{,}
\end{equation}
where $\Psi = 1$, ${{\mathbf{v}}^{*}}$, ${{\mathbf{v}}^{*}}\cdot {{\mathbf{v}}^{*}}+{{\eta }^{2}}$, ${{\mathbf{v}}^{*}}\cdot {{\mathbf{v}}^{*}}$, $\left( {{\mathbf{v}}^{*}}\cdot {{\mathbf{v}}^{*}}+{{\eta }^{2}} \right){{\mathbf{v}}^{*}}$, ${{\mathbf{v}}^{*}}{{\mathbf{v}}^{*}}{{\mathbf{v}}^{*}}$, $\left( {{\mathbf{v}}^{*}}\cdot {{\mathbf{v}}^{*}}+{{\eta }^{2}} \right){{\mathbf{v}}^{*}}{{\mathbf{v}}^{*}}$, 
correspondingly, 
${{\Psi }_{i}}=1$, $\mathbf{v}_{i}^{*}$, $\mathbf{v}_{i}^{*}\cdot \mathbf{v}_{i}^{*}+\eta _{i}^{2}$, $\mathbf{v}_{i}^{*}\cdot \mathbf{v}_{i}^{*}$, $\left( \mathbf{v}_{i}^{*}\cdot \mathbf{v}_{i}^{*}+\eta _{i}^{2} \right)\mathbf{v}_{i}^{*}$, $\mathbf{v}_{i}^{*}\mathbf{v}_{i}^{*}\mathbf{v}_{i}^{*}$, $\left( \mathbf{v}_{i}^{*}\cdot \mathbf{v}_{i}^{*}+\eta _{i}^{2} \right)\mathbf{v}_{i}^{*}\mathbf{v}_{i}^{*}$. Here ${{\mathbf{v}}^{*}}=\mathbf{v}-\mathbf{u}$ and $\mathbf{v}_{i}^{*}={{\mathbf{v}}_{i}}-\mathbf{u}$, with $\mathbf{u}$ the flow velocity. In addition, the equilibrium distribution function $f^{eq}$ is 
\begin{equation}
	{{f}^{eq}}=n{{\left( \frac{m}{2\pi T} \right)}^{D/2}}{{\left( \frac{m}{2\pi IT} \right)}^{1/2}}\exp \left[ -\frac{m{{\left| \mathbf{v}-\mathbf{u} \right|}^{2}}}{2T}-\frac{m{{\eta }^{2}}}{2IT} \right]
	\label{equilibrium function}
	\tt{,}
\end{equation}
where $D$ denotes the dimensional translational degree of freedom, $I$ stands for extra degrees of freedom due to vibration and/or rotation, and $\eta$ is used to describe the corresponding vibrational and/or rotational energies. The other parameters include $n$ as the particle number density, $T$ as the temperature, $m$ = 1 as the particle mass, and $\rho = nm$ as the mass density. Besides,  the specific heat ratio is $\gamma =\left( D+I+2 \right)/\left( D+I \right)$, the dynamic viscosity $ \mu={\rho T}{\tau} $, kinematic viscosity $\nu={\mu }/{\rho }={T}{\tau}$, and bulk viscosity ${{\mu }_{B}}=\mu \left( {2}/{D}-{2}/({D+I}) \right)$.

The central kinetic moments can be expressed in a unified form
\begin{equation}
	{{\bf{\bar f}}^{{\bf{eq}}}} = {\bf{M}}{{\bf{f}}^{{\bf{eq}}}}
	\label{matrix inversion}
	\tt{.}
\end{equation}
Here, ${{\bf{\bar f}}^{{\bf{eq}}}}={\left( {\begin{matrix}{\bar f_1^{eq}} \: {\bar f_2^{eq}} \: \cdots \: {\bar f_{16}^{eq}}\end{matrix}} \right)^{\rm{T}}}$ and ${{\bf{f}}^{eq}} = {\left( {\begin{matrix} {f_1^{eq}} \: {f_2^{eq}} \: \cdots \: {f_{16}^{eq}}\end{matrix}} \right)^{\rm{T}}}$ represent the equilibrium velocity distribution functions in the central moment space and velocity space, respectively.
The matrix ${\bf{M}}={{\left( {{\mathbf{M}}_{1}}\;{{\mathbf{M}}_{2}}\;\cdots \;{{\mathbf{M}}_{16}} \right)}^{\mathrm{T}}}$ acts as the bridge for transforming the velocity distribution function between the moment space and the discrete velocity space, which contains the blocks, ${{\mathbf{M}}_{i}}=\left( \begin{matrix} {{M}_{i1}} \: {{M}_{i2}} \: \cdots \:  {{M}_{i16}} \end{matrix} \right)$, 
with elements 
${M}_{1i}=1$, ${M}_{2i}=v_{ix}^{*}$, ${M}_{3i}=v_{iy}^{*}$, ${M}_{4i}=v_{i}^{{*}2}+\eta _{i}^{2}$, ${M}_{5i}=v_{ix}^{{*}2}$, ${M}_{6i}=v_{ix}^{*}v_{iy}^{*}$, ${M}_{7i}=$ $v_{iy}^{{*}2}$, ${M}_{8i}=(v_{i}^{{*}2}+\eta
_{i}^{2})v_{ix}^{*}$, ${M}_{9i}=(v_{i}^{{*}2}+\eta _{i}^{2})v_{iy}^{*}$, $%
{M}_{10i}=v_{ix}^{{*}3}$, ${M}_{11i}=v_{ix}^{{*}2}v_{iy}^{*}$, ${M}_{12i}=v_{ix}^{*}v_{iy}^{{*}2}$,
${M}_{13i}=v_{iy}^{{*}3}$, ${M}_{14i}=(v_{i}^{{*}2}+\eta _{i}^{2})v_{ix}^{{*}2}$, $%
{M}_{15i}=(v_{i}^{{*}2}+\eta _{i}^{2})v_{ix}^{*}v_{iy}^{*}$, ${M}_{16i}=(v_{i}^{{*}2}+\eta
_{i}^{2})v_{iy}^{{*}2}$. 
Consequently, Eq. \ref{matrix inversion} can be expressed as 
\begin{equation}
	{\bf{f}}^{{\bf{eq}}} = {\bf{M}}^{{\bf{ - 1}}}{\bf{\bar f}}^{{\bf{eq}}}
	\label{matrix inversion2}
	\tt{.}
\end{equation}

It should be mentioned that the mass, momentum and energy conservation are described by the first three kinetic moments in Eq. (\ref{moment}), where $f_i^{eq}$ can be replaced by $f_i$. In other words, the density, flow velocity and temperature are obtained from the kinetic moments of $f_i$. Replacing $f_i^{eq}$ with $f_i$ results in the imbalance in the last four kinetic moments. The difference between the results calculated by $f_i^{eq}$ and $f_i$ can be used to describe the deviation of the system from equilibrium state. Consequently, the CDBM contains the following nonequilibrium manifestations:
\begin{equation}
	{\mathbf \Delta}_{2}^{*}=\sum_{i}\Big(f_{i}-f_{i}^{eq}\Big)\mathbf{v}_{i}^{*}\mathbf{v}_{i}^{*}
	\label{delta2}
	\tt{,}
\end{equation}
\begin{equation}
	{\mathbf \Delta}_{3,1}^{*}=\sum_{i}\Big(f_{i}-f_{i}^{eq}\Big)\Big(\mathbf{v}_{i}^{*}\mathbf{v}_{i}^{*}+\eta_{i}^{2}\Big)\mathbf{v}_{i}^{*}
	\label{delta3,1}
	\tt{,}
\end{equation}
\begin{equation}
	{\mathbf \Delta }_{3}^{*}=\sum\nolimits_{i}{\left( {{f}_{i}}-f_{i}^{eq} \right)}\mathbf{v}_{i}^{*}\mathbf{v}_{i}^{*}\mathbf{v}_{i}^{*}
	\label{delta3}
	\tt{,}
\end{equation}
\begin{equation}
	{\mathbf \Delta}_{4,2}^{*}=\sum\nolimits_{i}{\left( {{f}_{i}}-f_{i}^{eq} \right)\left( \mathbf{v}_{i}^{*}\mathbf{v}_{i}^{*}+\eta _{i}^{2} \right)}\mathbf{v}_{i}^{*}\mathbf{v}_{i}^{*}
	\label{delta4,2}
	\tt{.}
\end{equation}
The second order tensor $\mathbf{\Delta} _{2}^{*}$ corresponds to the viscous stress tensor and twice the nonorganized energy. The vector $\mathbf{\Delta }_{3,1}^{*}$ is associated with the heat flux and twice the nonorganized energy flux. $\mathbf{\Delta} _{3}^{*}$ and $\mathbf{\Delta} _{4,2}^{*}$ are higher-order nonequilibrium quantities beyond traditional NS models \cite{ZYD2016CNF}. It is crucial to note that, in comparison with previous DBMs constructed in raw-moment space, the CDBM can provide the nonequilibrium effects related to the thermal fluctuation directly.

In addition, in the basic lattice Boltzmann model, spatial and temporal discretizations are coupled, which restricts the choice of discrete velocities and also affects the construction of the discrete equilibrium distribution function. By contrast, the discrete Boltzmann method retains the use of discrete velocities but eliminates dependence on specific discretization schemes. Instead, it directly solves the continuous Boltzmann equation, allowing the CDBM to adopt spatial and temporal discretization schemes flexibly. In the subsequent simulations, the temporal derivatives are computed using the second-order Runge-Kutta scheme \cite{LCD2022ACTA}, while the spatial discretization employs the second-order non-oscillatory, non-free-parameter dissipation finite difference (NND) scheme \cite{Zhang1991NND}. The details are as follows,
\begin{equation}
	{{v}_{ir}}\frac{\partial {{f}_{i}}}{\partial r}=-\frac{1}{\Delta r}[H(ir+\frac{1}{2})-H(ir-\frac{1}{2})]
	\label{governing equation1}
	\tt{,}
\end{equation}
\begin{equation}
	H(ir+\frac{1}{2})={{H}_{L}}(ir+\frac{1}{2})+{{H}_{R}}(ir+\frac{1}{2})
	\label{governing equation2}
	\tt{,}
\end{equation}
\begin{equation}
	{{H}_{L}}(ir+\frac{1}{2})=f_{i}^{+}(ir)+  
	\frac{1}{2}{\rm{minmod}}[\Delta f_{i}^{+}(ir+\frac{1}{2}),\Delta f_{i}^{+}(ir-\frac{1}{2})] \\ 
	\label{governing equation3}
	\tt{,}
\end{equation}
\begin{equation}
	{{H}_{R}}(ir+\frac{1}{2})=f_{i}^{-}(ir+1)-\frac{1}{2}{\rm{minmod}}[\Delta f_{i}^{-}(ir+\frac{1}{2}),\Delta f_{i}^{-}(ir+\frac{3}{2})]
	\label{governing equation4}
	\tt{.}
\end{equation}
The function ${\rm{minmod}}$ is also a type of flux limiter, defined as follows: 
\begin{equation}
	{\rm{minmod}}\left[X,Y\right]=\begin{cases}0,&Y=0\quad {\rm{or}}\quad XY\leqslant0\\[1ex]X,&\left|\frac{X}{Y}\right|\leqslant1\\[1ex]Y,&\left|\frac{X}{Y}\right|>1\end{cases}
	\label{minmod}
	\tt{.}
\end{equation}
In addition,
\begin{equation}
	\Delta f_{i}^{+}(ir+\frac{1}{2})=f_{i}^{+}(ir+1)-f_{i}^{+}(ir)
	\label{governing equation5}
	\tt{,}
\end{equation}
\begin{equation}
	\Delta f_{i}^{-}(ir+\frac{1}{2})=f_{i}^{-}(ir+1)-f_{i}^{-}(ir)
	\label{governing equation6}
	\tt{,}
\end{equation}
\begin{equation}
	f_{i}^{+}(ir)=\frac{1}{2}({{v}_{ir}}+\left| {{v}_{ir}} \right|){{f}_{i}}
	\label{governing equation7}
	\tt{,}
\end{equation}
\begin{equation}
	f_{i}^{-}(ir)=\frac{1}{2}({{v}_{ir}}-\left| {{v}_{ir}} \right|){{f}_{i}}
	\label{governing equation8}
	\tt{,}
\end{equation}
where $ir$ denotes the $i$-th grid point in the $r$ direction. The NND scheme achieves second-order spatial accuracy. This scheme can suppress odd-even decoupling oscillations and effectively capture strong discontinuities.

\section{Numerical validation}\label{SecIII}

In this section, let us verify that the CDBM is not only applicable for high-speed compressible flows, but can also capture the nonequilibrium effects accurately. To this end, five representative benchmarks are considered: the Sod shock tube and Lax shock tube validate the ability of the CDBM to capture discontinuities under varying specific heat ratios, the shock wave test demonstrates the suitability of the CDBM for hypervelocity compressible flows, the two-dimensional sound wave and decaying Taylor-Green vortex flow are used to verify the efficiency of CDBM in two-dimensional cases.

\subsection{Sod shock tube}

First, we consider two typical Riemann problems, i.e., the Sod shock tube and the Lax shock tube. For the Sod shock tube, the initial conditions are given by 
\begin{equation}
	\left\{ \begin{array}{*{35}{l}}
		{{\left(\rho, {u}_{x}, {u}_{y}, T \right)}_{L}} = \left( 1, 0, 0, 1 \right) \tt{,} \\
		{{\left(\rho, {u}_{x}, {u}_{y}, T \right)}_{R}} = \left( 0.125, 0, 0, 0.8 \right)
		\nonumber
		\tt{.} 
	\end{array} \right.
\end{equation}
Here, $L$ and $R$ denote the regions $0 < x \le 0.075$ and $0.075 < x \le 0.15$, respectively. The simulation parameters are specified as follows: the grid mesh is $N_x \times N_y = 1500 \times 1$, with a spatial resolution of $\Delta x = \Delta y = 1 \times 10^{-4}$, a time step of $\Delta t = 2 \times 10^{-5}$, and a specific heat ratio $\gamma = 2$. The discrete velocity set is $(v_a, v_b, v_c, v_d) = (2.5, 2.5, 1.2, 1.1)$ and $(\eta_a, \eta_b, \eta_c, \eta_d) = (0, 0, 0, 1.2)$. Additionally, the relaxation time is set to $\tau=1 \times 10^{-4}$.
\begin{figure}
	\begin{center}
		\includegraphics[width=0.95\textwidth]{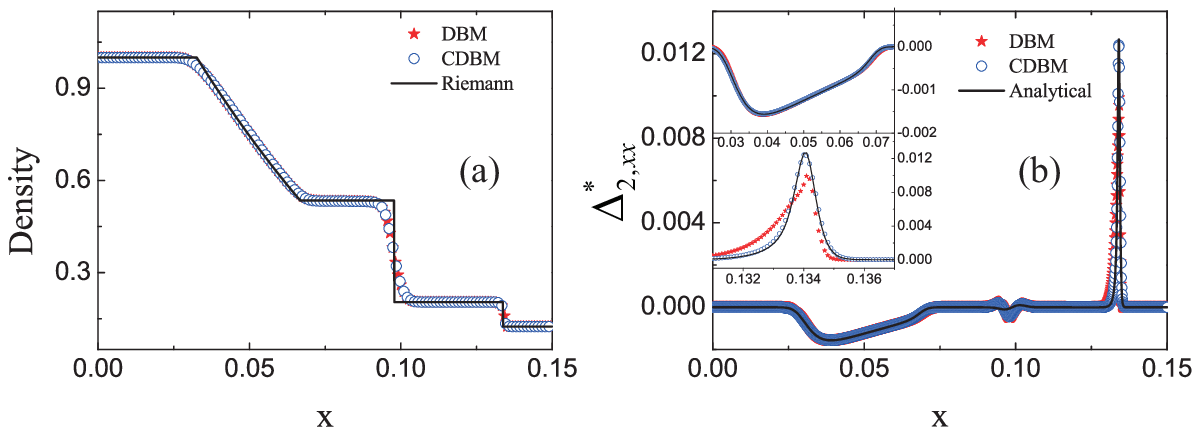}
	\end{center}
	\caption{Profiles of the density ${\rho}$ (a) and the nonequilibrium quantity ${\Delta} _{2,xx}^{*}$ (b) in the Sod shock tube.}
	\label{Fig02}
\end{figure}
Figure \ref{Fig02} (a) illustrates the evolution of density in the Sod shock tube. The stars, circles, and solid line correspond to the DBM results, CDBM results, and the Riemann solutions, respectively. The simulation results exhibit good agreement with the analytical Riemann solutions. Some discrepancies are observed between the numerical results and the Riemann solutions, particularly near the rarefaction wave, material interface, and shock front. These arise from the fact that the CDBM simulation results incorporate essential thermodynamic nonequilibrium effects, which are neglected by the Riemann solutions. 

Figure \ref{Fig02} (b) shows the profile of twice the nonorganized energy along the $x$ direction. The stars and circles denote the simulation results obtained from the DBM and CDBM, while the solid line represents the exact solution \cite{Lin2019PRE} 
\begin{equation}
	\Delta _{2,xx}^{*}={2\rho T}{\tau}\biggl(\frac{1-D-I}{D+I}\frac{\partial u_x}{\partial x}+\frac{1}{D+I}\frac{\partial u_y}{\partial y}\biggr)
	\label{delta2,xx}
	\tt{.}
\end{equation}
Obviously, the nonorganized energy near rarefaction wave exhibits negativity which is illustrated in the upper subplot, and a peak emerges around the shock wave as depicted in the lower subplot. Furthermore, the nonorganized energy approaches zero in other regions, with minor numerical oscillations observed at the contact wave. In fact, the minor undershoot observed in the Sod shock tube arises due to the sharp discontinuity in the physical field inherent in the initial configuration. This artificial discontinuity deviates from the characteristics of a natural interface, which typically exhibits a smooth transition layer. Consequently, a minor undershoot develops near this discontinuity during the early stages and gradually dissipates over time. Importantly, the CDBM results align more closely with the exact solution than the DBM results across the entire profile, especially in the peak region. This provides evidence that the CDBM is highly effective in capturing nonequilibrium effects in various regions of the Sod shock tube.

\subsection{Lax shock tube}

The initial condition for the Lax shock tube is 
\begin{equation}
	\left\{ \begin{array}{*{35}{l}}
		{{\left(\rho, {u}_{x}, {u}_{y}, T \right)}_{L}} = \left( 0.445, 0.698, 0, 7.928\right) \tt{,} \\
		{{\left(\rho, {u}_{x}, {u}_{y}, T \right)}_{R}} = \left( 0.5, 0, 0, 1.142 \right)
		\nonumber
		\tt{,} 
	\end{array} \right.
\end{equation}
where $L \in [0,1)$ and $R \in [1,2]$ stand for the left and right sides, respectively. The computational grid is specified as $N_x \times N_y = 1000 \times 1$, with a spatial resolution of $\Delta x = \Delta y = 2 \times 10^{-3}$ and a time step of $\Delta t = 2 \times 10^{-5}$. In addition, the specific heat ratio is $\gamma = 1.4$, and the discrete velocity set is $(v_a, v_b, v_c, v_d) = (4.4, 4.4, 3, 1.8)$ and $(\eta_a, \eta_b, \eta_c, \eta_d) = (0, 0, 5, 0)$. The relaxation time is $\tau=1 \times 10^{-4}$.
\begin{figure}
	\begin{center}
		\includegraphics[width=0.8\textwidth]{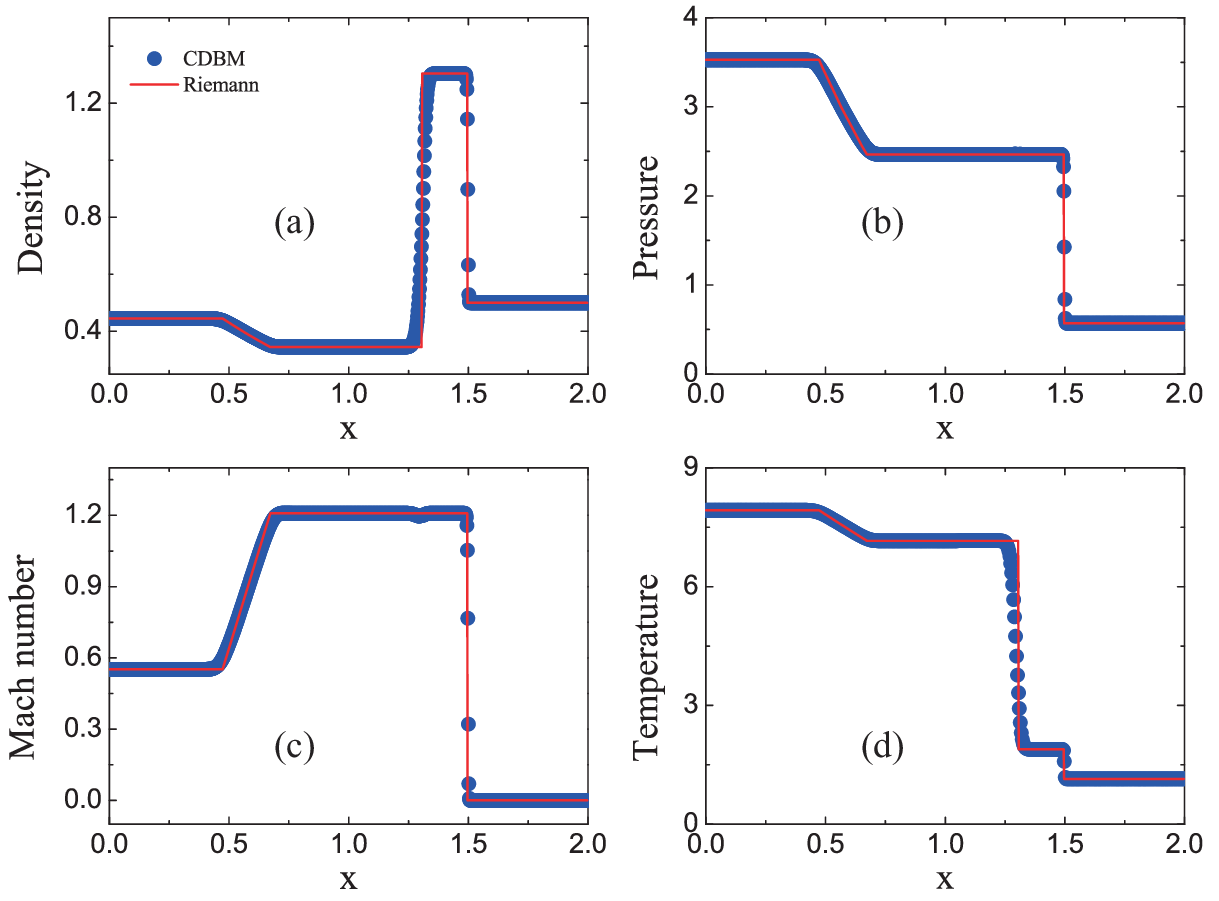}
	\end{center}
	\caption{Profiles of the density ${\rho}$ (a), temperature $T$ (b), Mach number $\rm{Ma}$ (c), and pressure $P$ (d) in the Lax shock tube.}
	\label{Fig03}
\end{figure}
Figure \ref{Fig03} presents the results of the Lax problem at $t = 0.2$. The Riemann solutions are represented by lines, while the symbols denote the computed quantities obtained using the CDBM. It can be observed that the simulation results exhibit a good agreement with the analytical solutions. To be specific, there is a rarefaction wave propagating to the left, a material interface in the center, and a shock wave traveling to the right. Consequently, the CDBM successfully captures all three distinct structures.

\subsection{Shock wave}

The CDBM is suitable for high-speed compressible flow. To assess its performance at high Mach numbers, we simulate a shock wave with a Mach number of $\rm{Ma} = 15$. The initial conditions are specified as follows, 
\begin{equation}
	\left\{ \begin{array}{*{35}{l}}
		{{\left(\rho, {u}_{x}, {u}_{y}, T, P \right)}_{L}} = \left( 5.8696, 14.7245, 0, 44.6938, 262.3333 \right) \tt{,} \\
		{{\left(\rho, {u}_{x}, {u}_{y}, T, P \right)}_{R}} = \left( 1, 0, 0, 1, 1 \right)
		\nonumber
		\tt{.} 
	\end{array} \right.
\end{equation}
Here, $L$ and $R$ stand for  $0 < x \le 0.02$ and $0.02 < x \le 1.5$, respectively. The simulation parameters are as follows: the grid mesh is $N_x \times N_y = 10000 \times 1$, the spatial resolution is $\Delta x = \Delta y = 2 \times 10^{-4}$, the time step is $\Delta t = 2 \times 10^{-6}$, the specific heat ratio is $\gamma = 1.4$. The discrete velocity set is given as $(v_a, v_b, v_c, v_d) = (21.8, 21.8, 17.8, 9.8)$ and $(\eta_a, \eta_b, \eta_c, \eta_d) = (18.8, 0, 0, 0)$.

\begin{figure}
	\begin{center}
		\includegraphics[width=0.88\textwidth]{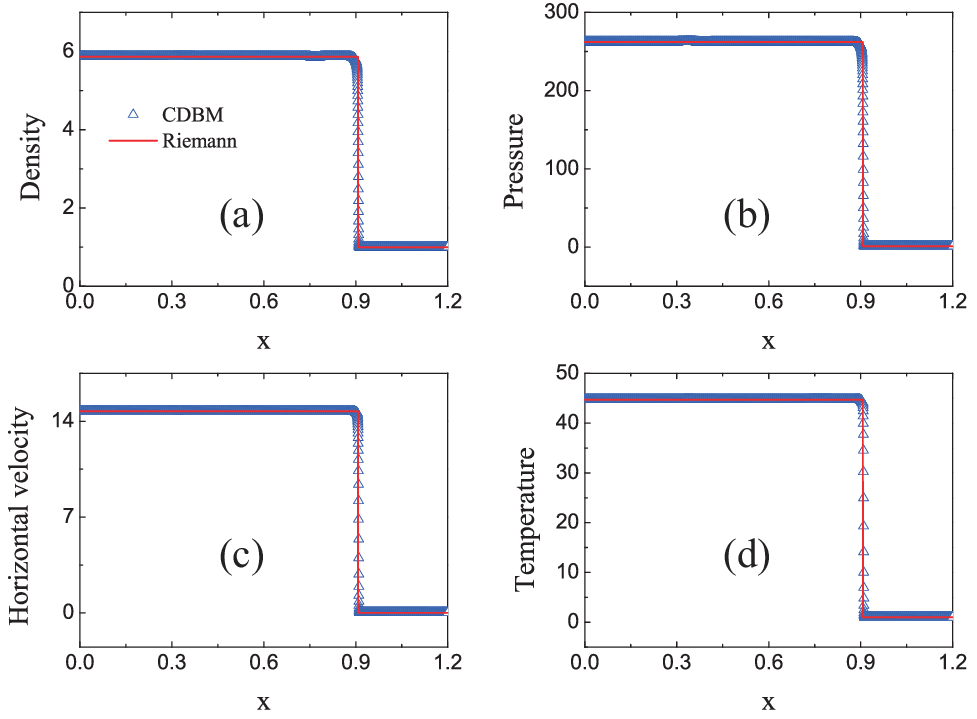}
	\end{center}
	\caption{Profiles of the density (a), pressure (b), horizontal velocity (c) and temperature (d) in the shock wave.}
	\label{Fig04}
\end{figure}

Figures \ref{Fig04} (a)-(d) illustrate the profiles of density, pressure, horizontal velocity and temperature, respectively. The symbols represent the CDBM results, The symbols represent the CDBM results, while the solid lines denote the Riemann solutions. Clearly, the CDBM results align closely with the Riemann solutions. Therefore, the CDBM successfully captures the shock wave at a Mach number of $\rm{Ma}=15$. In other words, the model is suitable for a wide range of flow regimes, from incompressible flows to hypervelocity compressible flows.

\subsection{Sound wave}

\begin{figure}
	\begin{center}
		\includegraphics[width=0.5\textwidth]{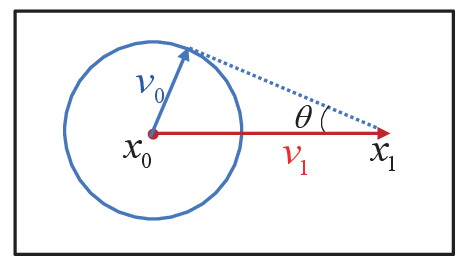}
	\end{center}
	\caption{The sketch of the propagation for the two-dimensional sound wave.}
	\label{Fig05}
\end{figure}

Then, the sound wave simulation verifies the suitability of the CDBM for compressible flows. Figure \ref{Fig05} illustrates the initial condition. A perturbation is introduced at position $x_0$, spreading at the speed of sound. Simultaneously, the perturbation propagates rightward at twice the speed of sound. Therefore, the angle of propagation can be utilized to verify the accuracy of our model.

The grid mesh size is $N_x \times N_y = 1600 \times 1200$, with a spatial resolution of $\Delta x = \Delta y = 5 \times 10^{-4}$, and a time step of $\Delta t = 2 \times 10^{-5}$. In addition, the relaxation time is $\tau=1 \times 10^{-4}$, the kinematic viscosity is $\mu=1 \times 10^{-4}$, the specific heat ratio is $\gamma = 1.4$, and the discrete velocity set is $(v_a, v_b, v_c, v_d) = (1.4, 1.1, 1.1, 1)$ and $(\eta_a, \eta_b, \eta_c, \eta_d) = (3, 0, 0, 0)$.

\begin{figure}
	\begin{center}
		\includegraphics[width=0.5\textwidth]{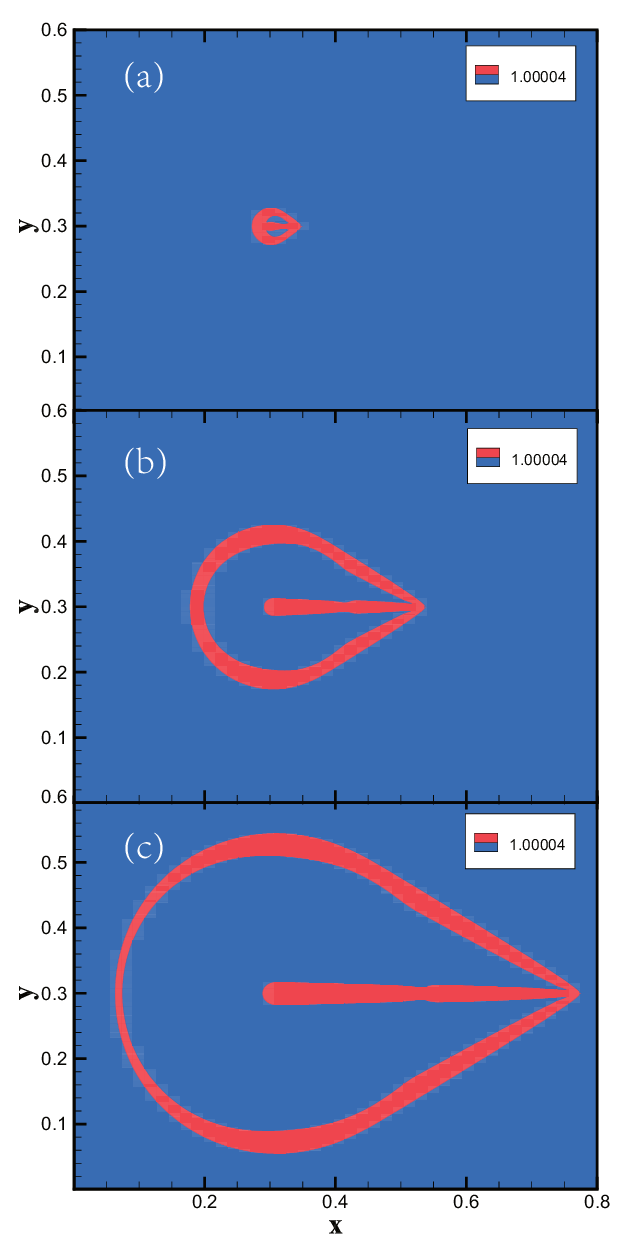}
	\end{center}
	\caption{Profiles of the two-dimensional sound waves at various time instants, (a) $t$ = 0.02, (b) $t$=0.1, and (c) $t$=0.2.}
	\label{Fig06}
\end{figure}

Figure \ref{Fig06} illustrates the propagation of two-dimensional sound waves at different times: (a) $t$ = 0.02, (b) $t$=0.1, and (c) $t$=0.2. At $t$=0.2, the exact solution is $\sin\theta = 0.5$, while the simulation result is $\sin\theta = 0.50106$. The relative error between the simulation result and the exact solution is $0.212\%$, indicating satisfactory agreement. These results demonstrate that the CDBM is well-suited for modeling two-dimensional compressible waves.

\subsection{Taylor-Green vortex flow}

Finanlly, the CDBM was validated by a two-dimensional Taylor-Green vortex flow. The solution of this flow problem can be given analytically as
\begin{equation}
	u_x(x,y,t)=-u_{0}\cos(\pi x/L)\sin(\pi y/L)e^{-2\pi^{2}u_{0}t/({\mathop{\rm Re}\nolimits}  L)}
	\label{ux}
	\tt{,}
\end{equation}
\begin{equation}
	u_y(x,y,t)=+u_{0}\sin(\pi x/L)\cos(\pi y/L)e^{-2\pi^{2}u_{0}t/({\mathop{\rm Re}\nolimits} L)}
	\label{uy}
	\tt{,}
\end{equation}
\begin{equation}
	p\bigl(x,y,t\bigr)=p_{0}-\frac{p_{0}u_{0}^{2}}{4}\Bigl[\cos\bigl(2\pi x/L\bigr)+\cos\bigl(2\pi y/L\bigr)\Bigr]e^{-4\pi^{2}u_{0}t/\bigl({\mathop{\rm Re}\nolimits}L\bigr)}
	\label{p}
	\tt{,}
\end{equation}
where $u_0=0.01$ denotes the reference velocity, $p_0=1.0$ represents the reference pressure, $L=0.05$ signifies the reference length, and the Reynolds number, ${\mathop{\rm Re}\nolimits}$, is defined as ${\mathop{\rm Re}\nolimits}=\rho_0u_0L/\mu = 1 $, where kinematic viscosity is $\mu=5 \times 10^{-4}$.
The computational domain for this flow problem extends over $[0,2L]$, with the grid discretization performed using a mesh of size $N_x \times N_y = 100 \times 100$. Additional parameters for this simulation include the spatial resolution $\Delta x = \Delta y = 1 \times {10^{ - 3}}$, the temporal step $\Delta t = 2 \times {10^{ - 5}}$, a specific heat ratio of $\gamma=1.4$, a discrete velocity set of $(v_a, v_b, v_c,v_d)=(0.2, 0.2, 2, 2)$, and $(\eta_a, \eta_b, \eta_c, \eta_d)= (0, 5, 0, 2.6)$.
Moreover, the relaxation time is $\tau=5 \times 10^{-4}$.
\begin{figure}
	\begin{center}
		\includegraphics[width=1\textwidth]{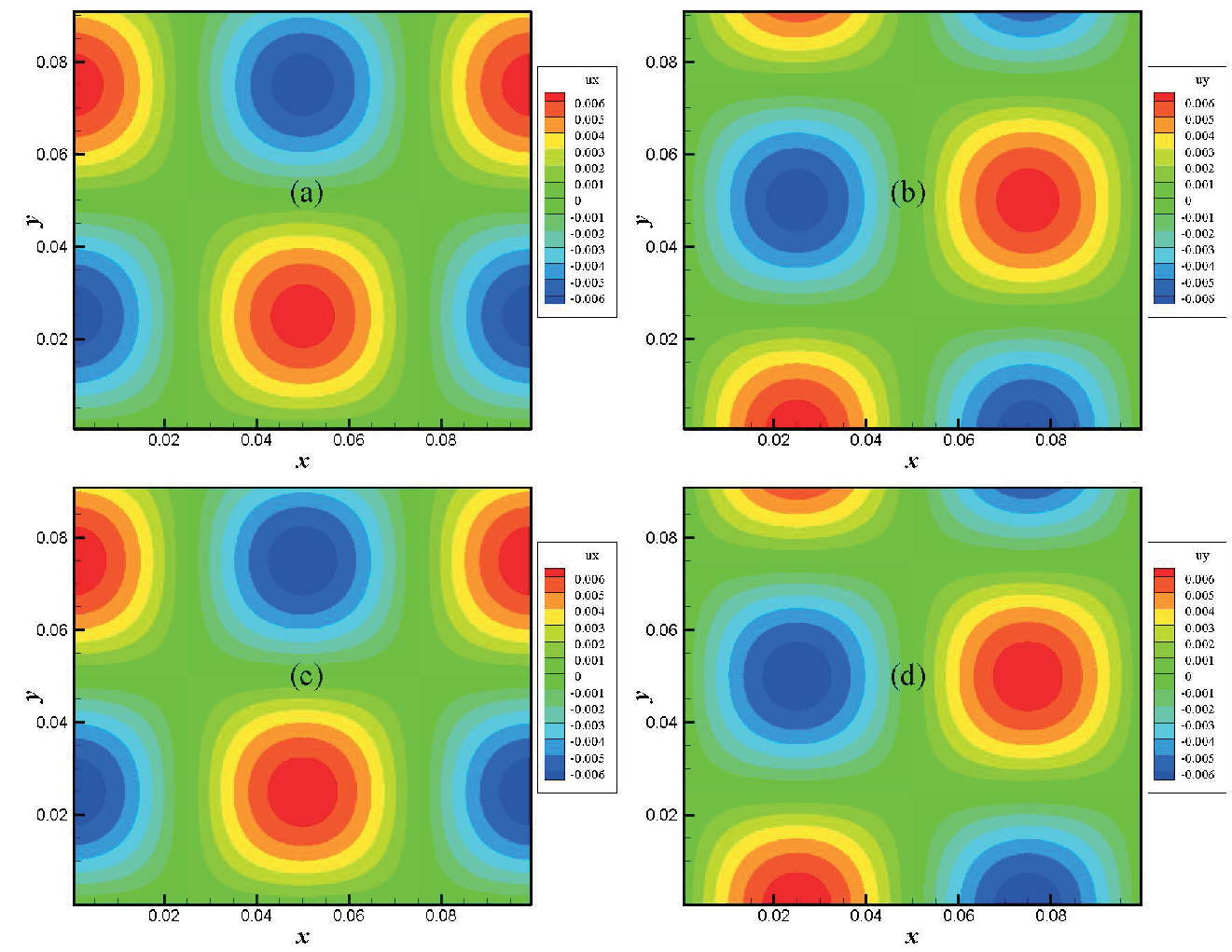}
	\end{center}
	\caption{Comparison of CDBM simulation and analytical solutions for Taylor-Green vortex velocity fields at $t=0.1$. (a) Horizontal velocity of CDBM, (b) vertical velocity of CDBM, (c) Horizontal velocity of analytical solution, and (d) vertical velocity of analytical solution.}
	\label{Fig07}
\end{figure}
In this section, we compare the simulation results obtained from the CDBM with the exact analytical solutions for the Taylor-Green vortex flow. Figures \ref{Fig07} (a) and (b) depict the horizontal and vertical velocity fields obtained from the CDBM simulation, whereas Figs. \ref{Fig07} (c) and (d) illustrate the corresponding analytical results. Both sets of results exhibit a high degree of qualitative agreement, demonstrating similar periodic structures and symmetry in the velocity fields. The close alignment of velocity magnitudes, as indicated by the color bars, further confirms the accuracy of the CDBM simulation.

In addition, the relative error of velocity component between the numerical result and the analytical solution is measured using the $L_2$ norm, defined as follows,
\begin{equation}
	{{L}_{2}}{{\left( \begin{matrix}
				{{u}_{x}}  \\
			\end{matrix} \right)}_{relative}}=\sqrt{\frac{1}{{{N}_{x}}\times {{N}_{y}}}\sum\limits_{i,j}{{{\left( \frac{{{u}_{x}}_{\left( i,j \right)}^{numerical}-{{u}_{x}}_{\left( i,j \right)}^{exact}}{{{u}_{0}}} \right)}^{2}}}}
	\label{L2}
	\tt{.}
\end{equation}
Five different uniform grid meshes with sizes of $20\times 20$, $30\times 30$, $40\times 40$, $50\times 50$, and $60\times 60$ are used to discretize the domain. The corresponding spatial steps are $5 \times 10^{-3}$, $3.34 \times 10^{-3}$, $2.5 \times 10^{-3}$, $2 \times 10^{-3}$, and $1.67 \times 10^{-3}$, respectively.
\begin{figure}
	\begin{center}
		\includegraphics[width=0.5\textwidth]{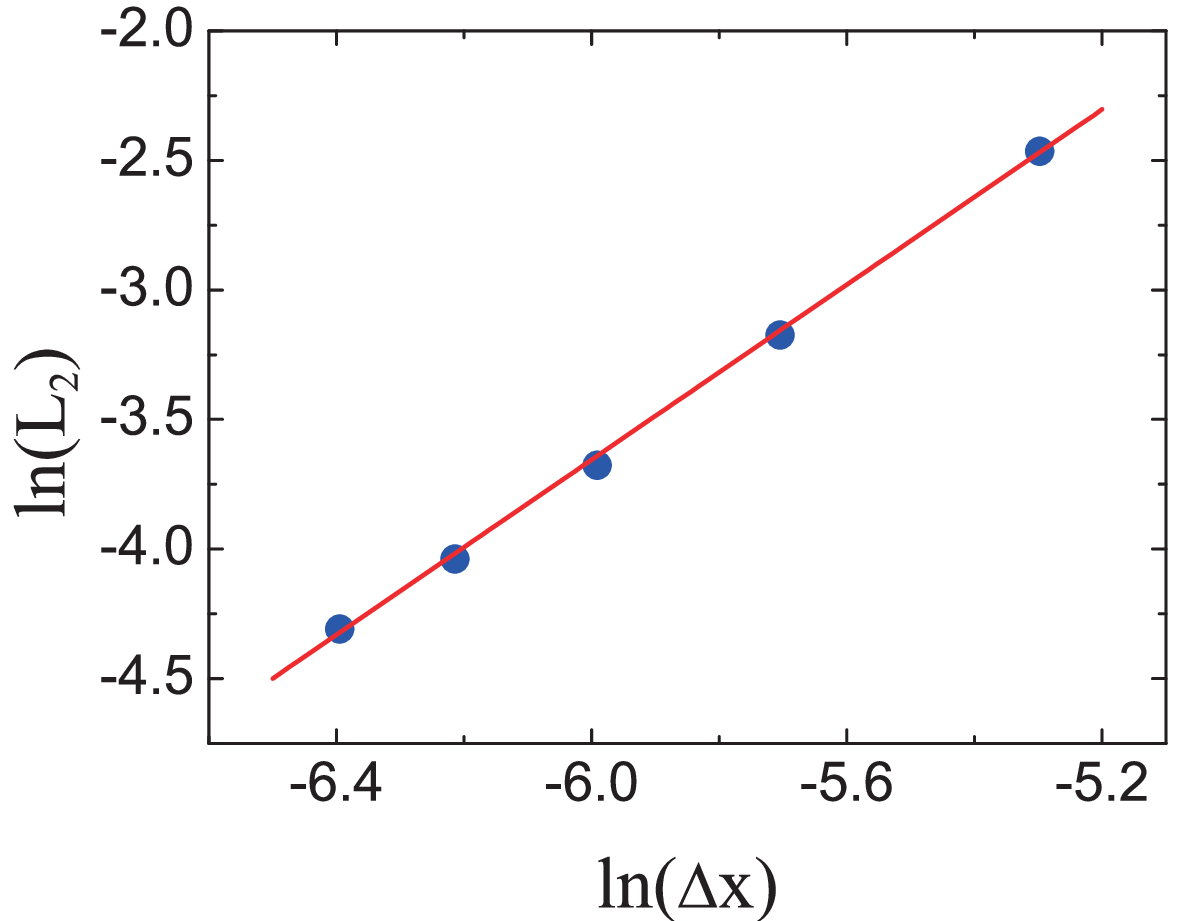}
	\end{center}
	\caption{$L_2$ norm of the relative error of horizontal velocity versus space step $\Delta x$ for decaying vortex flow.}
	\label{Fig08}
\end{figure}
Figure \ref{Fig08} showes the numerical results of $L_2$ norms, where the space step $\Delta x$ is plotted on a logarithmic scale. The slope of the line is 1.69, which is near the second order in theory. These findings suggest that the CDBM method effectively captures the dynamics of the Taylor-Green vortex flow, validating its reliability for studying complex two-dimensional fluid flow phenomena.

\section{Conclusion}\label{SecV}

In summary, a CDBM is developed for kinetic modeling of compressible flows with both hydrodynamic and thermodynamic nonequilibrium effects. The central kinetic moments are employed for calculating the equilibrium discrete distribution functions. Conservation moments of the discrete velocity distribution function are utilized to derive macroscopic physical quantities, while higher-order central kinetic moments are employed to characterize nonequilibrium effects. Its ability to capture nonequilibrium effects is demonstrated through the simulation of the Sod shock tube. The results exhibit excellent agreement with exact solutions. Furthermore, the Lax shock tube benchmark confirms that the CDBM can accurately capture discontinuities under different specific heat ratios. Additionally, the simulation of shock waves at Mach number $\rm{Ma} = 15$ verifies the capability of the CDBM for hypervelocity compressible flows. The results of two-dimensional sound wave propagation and decaying Taylor-Green vortex flow further showcase the efficiency of the CDBM in two-dimensional cases. Overall, the proposed CDBM proves to be a robust and promising tool for investigating complex compressible fluid systems, particularly those with significant thermodynamic nonequilibrium effects at the mesoscopic level.

\section*{Acknowledgements}

This work is supported by Guangdong Basic and Applied Basic Research Foundation (under Grant No. 2022A1515012116), China Scholarship Council (Nos. 202306380179 and 202306380288), and Fundamental Research Funds for the Central Universities, Sun Yat-sen University (under Grant No. 24qnpy044). Support from the UK Engineering and Physical Sciences Research Council under the project “UK Consortium on Mesoscale Engineering Sciences (UKCOMES)” (Grant No. EP/X035875/1) is gratefully acknowledged. This work made use of computational support by CoSeC, the Computational Science Centre for Research Communities, through UKCOMES.

\bibliographystyle{tfq}
\bibliography{Reference}


\end{document}